\documentclass[twocolumn,pra,showpacs]{revtex4}
\usepackage{amssymb}
\usepackage{amsmath}
\usepackage{graphicx}
\begin{document}
\title{Quantum logic gates with controllable and selective interaction for superconducting charge qubits via a nanomechanical
resonator}
\author{Jie-Qiao Liao, Qin-Qin Wu and  Le-Man Kuang\footnote{To whom correspondence should be addressed.
Email: lmkuang@hunnu.edu.cn (L. M. Kuang)}} \affiliation{Key
Laboratory of Low-Dimensional Quantum Structures and Quantum
Control of Ministry of Education, and Department of Physics, Hunan
Normal University, Changsha 410081, People's Republic of China}
\begin{abstract}
In this paper, we propose a scheme to implement two-qubit logic
gates with a controllable and selective interaction in a scalable
superconducting circuit of charge qubits. A nanomechanical
resonator is used as a data bus to connect qubits. It is indicated
that a controllable interaction between qubits can be obtained by
making use of the data bus. It is shown that a selective
interaction between qubits can be realized when many qubits are
involved in the system under our consideration.
\end{abstract}

 \maketitle

Recently, much attention has been paid to superconducting quantum
circuits (SQCs) due to their potential applications in quantum
information processing \cite{Mak,Wen,You}. In SQCs there are
usually three types of qubits, i.e.,  charge qubits
\cite{Shn,Nak,Pas,Yam,Vio}, phase qubits \cite{Mar,Coo,Yu,Ber,Mc},
and flux qubits \cite{Orl,Moo,Wal,Fri,Eve,Mig}. As solid systems,
SQCs have the advantage of being able to be integrated to a large
scale.

In order to realize quantum computation, one need a universal
quantum logic gate set which consists of single-qubit rotation
operations and a two-qubit logic gate \cite{Nie}. Single-qubit
rotation operations are easy to be realized while two-qubit logic
gates are complex to be implemented since it is difficult to
obtain interaction between two qubits. Consequently, how to
implement a two-qubit logic gate becomes an important and
challenging topic in implementing quantum computation. Currently,
for superconducting systems there are two conceptually different
methods to obtain interaction between qubits. One is to employ
direct interaction between two qubits through connecting them with
capacitors, Josephson junctions, and dc-SQUID
\cite{Wen,You,Ave,Bla1,Joh}. In this method, we only obtain the
nearest neighbor interaction between qubits. Thus we can not
couple selectively any two qubits in a large qubit network. The
other method is to obtain effective interaction between two qubits
by coupling them to a boson mode called data bus
\cite{Cir,Cir2,Mon,Wall,Bla2}. The second method can selectively
couple any two qubits in a large qubit network. In practice,
selective and controllable interaction between two qubits is
expected in implementing quantum computation \cite{Mak1,Liu}.
Recently, some experiments have demonstrated interaction between
two superconducting qubits \cite{Pla,Nis,Him,Plo,Har}. It should
be mentioned that, however, in these experiments the interactions
between two qubits are only controllable but not selective. In
this letter, we propose a SQC scheme to implement two-qubit logic
gates with a controllable and selective interaction. In our
scheme, we introduce a new type of data bus: a nanomechanical
resonator (NAMR) \cite{Ble,Sch}. We show that a controllable
interaction between qubits can be obtained by making use of the
data bus, and a selective interaction between qubits can be
realized when many qubits are involved.

The system under our consideration consists of two superconducting
cooper-pair box (CPB) qubits fabricated by inserting a
superconducting loop by two identical Josephson junctions as shown
in Fig. $1$. The two superconducting loops share a common circuit
in which an NAMR is built. Two gate voltage sources $V_{g_1}$ and
$V_{g_2}$ are used to control the two CPBs through corresponding
gate capacitors. Moreover, the two CPBs can also be manipulated
via external magnetic fluxes threading the loops. The physical
mechanism of the interaction between the two CPBs can be explained
as follows. The vibration of the NAMR changes the effective areas
of  two loops and the magnetic fluxes in the loops \cite{Zho,Buk}.
When the area of the first loop increases (decreases),  the area
of the second loop decreases (increases). This correlative
relation between the areas of two loops leads to an effective
interaction between two qubits.

\begin{figure}[htp]
\center
\includegraphics[scale=0.8]{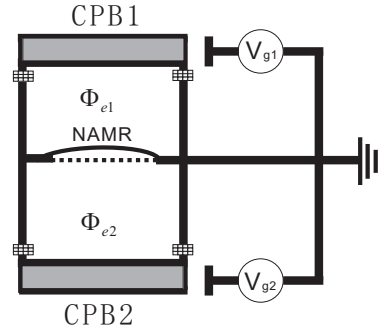}
\caption{ Schematic diagram of the system of two CPBs coupled with
an NAMR. Here $V_{gk}$ and $\Phi_{ek}$ are  gate voltages and
external biasing fluxes, respectively.}
\end{figure}

The Hamiltonian of the total system \cite{Mak} can be written as
\begin{eqnarray}
\label{e1}
H=&&\omega b^{\dag}b+\sum_{k=1}^{2}2E_{Ck}(2N_{gk}-1)\sigma_{zk}\nonumber\\
&&-\sum_{k=1}^{2}E_{Jk}\cos\left(\pi\frac{\Phi_{ek}}{\Phi_{0}}\right)\sigma_{xk},
\end{eqnarray}
where $\omega$ is the  flexural-mode frequency of the NAMR denoted
by operators $b$ and $b^{\dag}$. $\Phi_{0}=h/2e$ and $\Phi_{ek}$
are the flux quanta and bias flux of $k$th CPB, respectively. The
single-electron charging energy is given by
$E_{ck}=e^{2}/[2(2C_{Jk}+C_{gk})]$  where $C_{Jk}$ and $C_{gk}$
are the capacitance of each Josephson junction and the gate
capacitance in the CPB $k$, respectively. The  gate charge number
is given by  $N_{gk}=C_{gk}V_{gk}/2e$ with $V_{gk}$ being gate
voltage of $k$th CPB. The gate charge number can be controlled
through the gate voltage. The Pauli operators used in Eq.
(\ref{e1}) are defined by
\begin{eqnarray}
\label{e2}
\sigma_{zk}=|0\rangle_{k}\langle0|_{k}-|1\rangle_{k}\langle1|_{k},\nonumber\\
\sigma_{xk}=|0\rangle_{k}\langle1|_{k}+|1\rangle_{k}\langle0|_{k},
\end{eqnarray}
where CPB's charge states $|0\rangle_{k}$ and $|1\rangle_{k}$
correspond to zero and one extra cooper-pair on the inland,
respectively. Consistent with Fig. $1$, we denote the upper and
nether CPBs with indexes $1$ and $2$, respectively. We set
$\hbar=1$ throughout this paper.

The external magnetic fluxes in $k$th loop \cite{Zho,Buk} can be
expressed as the sum of two terms,
\begin{eqnarray}
\label{e3} \Phi_{ek}=\Phi_{bk}+(-1)^{k}BLx,
\end{eqnarray}
where the first term $\Phi_{bk}$ is the external magnetic fluxes
in loop $k$ when the NAMR does not vibrate while the second term
describes contribution from the vibration of the NAMR with $B$
being the magnetic field biasing on the two loops,  $L$  the
effective length, and $x$ the displacement of the NAMR. Assume
that the mass of the NAMR is $m$ and then we can write the
displacement as $x=(b^{\dag}+b)/\sqrt{2m\omega}$.

By substituting Eq. (\ref{e3}) into Eq. (\ref{e1}), we reduce the
expression of Eq. (\ref{e1}) to the following form,
\begin{eqnarray}
\label{e4} H=&&\omega b^{\dag}b+\sum_{k=1}^{2}2E_{ck}(2N_{gk}-1)\sigma_{zk}\nonumber\\
&&-\sum_{k=1}^{2}E_{Jk}\left[\cos\left(\pi\frac{\Phi_{bk}}{\Phi_{0}}\right)\cos\left(\pi\frac{BLx}{\Phi_{0}}\right)\right.\nonumber\\
&&\left.-(-1)^{k}\sin\left(\pi\frac{\Phi_{bk}}{\Phi_{0}}\right)\sin\left(\pi\frac{BLx}{\Phi_{0}}\right)\right]\sigma_{xk}.
\end{eqnarray}

It is straightforward to see that by controlling the biasing
fluxes $\Phi_{bk}$ we can choose the sine or cosine parts in the
Hamiltonian given by Eq. (\ref{e4}). In particular, when
$\sin(\pi\Phi_{bk}/\Phi_{0})=1$, the Hamiltonian given by Eq.
(\ref{e4}) becomes
\begin{eqnarray}
\label{e5} H=&&\omega b^{\dag}b+\sum_{i=1}^{2}2E_{Ck}(2N_{gk}-1)\sigma_{zk}\nonumber\\
&&+\sum_{k=1}^{2}(-1)^{k}E_{Jk}\sin\left(\pi\frac{BLx}{\Phi_{0}}\right)\sigma_{xk},
\end{eqnarray}
which can be reduced to the following form by expanding the sine
function up to first-order in $x$,
\begin{eqnarray}
\label{e6} H=\omega
b^{\dag}b+\sum_{k=1}^{2}\left[\frac{\omega_{k}}{2}\sigma_{zk}+(-1)^{k}g_{k}(b^{\dag}+b)\sigma_{xk}\right],
\end{eqnarray}
where the effective energy separation and the coupling constant
are given by
\begin{eqnarray}
\label{e6-1} \omega_{k}=4E_{Ck}(2N_{gk}-1), \hspace{0.3cm}
g_{k}=E_{Jk}\pi BL/(\Phi_{0}\sqrt{2m\omega}).
\end{eqnarray}

The Hamiltonian (\ref{e6}) is well known in quantum optics since
it is the same as the Hamiltonian of two two-level atoms
interacting with a cavity field. Differently, the effective atomic
energy separation in the present model can be controlled by tuning
the gate voltage. For coherent manipulation of two interacting
systems, one hopes to obtain a controllable interaction between
the two systems. However, the coupling constant $g_{k}$ in Eq.
(\ref{e6}) is fixed for a given system, we can not turn on or off
the coupling between the two CPBs and the NAMR on demand. In our
present system, fortunately, we can obtain a controllable coupling
between the two CPBs and the NAMR by replacing every Josephson
junction in Fig. $1$ by a SQUID-based superconducting loop
\cite{Mak}. Every loop contains two Josephson junctions with the
same Josephson energy $E^{0}_{Jk}$. For $k$th CPB, there are three
loops, left small one, right small one and middle big one. We
assume that the fluxes $\Phi_{lk}$ and $\Phi_{rk}$ which bias
respectively on left and right small loops have the same magnitude
but opposite sign, i.e., $\Phi_{lk}=-\Phi_{rk}=\Phi_{xk}$. The
coupling constant between $k$th CPB and the NAMR becomes
\begin{eqnarray}
\label{e6-2} g'_{k}=2E^{0}_{Jk}\cos(\pi\Phi_{xk}/\Phi_{0})\pi
BL/(\Phi_{0}\sqrt{2m\omega}),
\end{eqnarray}
which implies that we can control the coupling between the CPBs
and the NAMR by tuning these biasing fluxes.

For realizing quantum computation, a set of universal quantum
logic gates is necessary. A set of universal quantum logic gates
consists of single-qubit logic operation and a nontrivial
two-qubit logic gate such as CNOT and CP gate. In what follows we
will show how to implement two-qubit logic gates between the two
CPBs. From the Hamiltonian (\ref{e6}), we control the external
fluxes and gate voltages such that only one $k$th CPB  couples to
the NAMR, the Hamiltonian becomes
\begin{eqnarray}
\label{e7} H_{I}=\omega
b^{\dag}b+(-1)^{k}g'_{k}(b+b^{\dag})\sigma_{xk}.
\end{eqnarray}

In the interaction picture with respect to $H_{0}=\omega
b^{\dag}b$, the Hamiltonian given by Eq. (\ref{e7}) is transformed
to the following expression,
\begin{eqnarray}
\label{e8} H'_{k}(t_{k})=(-1)^{k}g'_{k}(be^{-i\omega
t_{k}}+b^{\dag}e^{i\omega t_{k}})\sigma_{xk},
\end{eqnarray}
which leads to a unitary evolution operator,
\begin{eqnarray}
\label{e9}U(t_{k})=\mathcal{\overleftarrow{T}}\exp\left[-i\int_{0}^{t_{k}}dsH'_{k}(s)\right],
\end{eqnarray}
where $\mathcal{\overleftarrow{T}}$ is the time-ordering operator.
Up to a trivial phase factor this unitary evolution operator can
be written as  \cite{Bre}
\begin{eqnarray}
\label{e10}V(\alpha_{k}\sigma_{xk})=\exp[b^{\dag}\alpha_{k}\sigma_{xk}-b\alpha^{\ast}_{k}\sigma_{xk}],
\end{eqnarray}
where we have introduced the coupling parameter between $k$th
qubit and the NAMR
\begin{equation}
\alpha_{k}=(-1)^{k}g'_{k}\left(1-e^{i\omega t_{k}}\right)/\omega,
\end{equation}
which can be controlled with
external flux $\Phi_{xk}$ and evolution time $t_{k}$.

In what follows we show how to realize two-qubit gate by using the
unitary evolution given by Eq. (\ref{e10}). We note that the
evolution operator given by Eq. (\ref{e10}) is a controlled
displacement operator \cite{Spi}, which can produce a displacement
of the NAMR, conditioned on eigenstates of the operator
$\sigma_{xk}$. It has been shown that, using the proper
controlled-displacement operators we can implement two-qubit
gates. The sequence of operations is arranged as follows,
\begin{eqnarray}
\label{e11}U(\alpha_{1},\alpha_{2})&=&V(\alpha_{2}\sigma_{x2})\otimes
V(\alpha_{1}\sigma_{x1})V(-\alpha_{2}\sigma_{x2})\nonumber\\
&&\otimes V(-\alpha_{1}\sigma_{x1}),
\end{eqnarray}
which can be written as
\begin{eqnarray}
\label{e12}U(\alpha_{1},\alpha_{2})=\exp[i\theta\sigma_{x1}\sigma_{x2}],
\end{eqnarray}
where the effective coupling constant
\begin{eqnarray}
\theta=2|\alpha_{1}||\alpha_{2}|\sin(\varphi_{2}-\varphi_{1})
\end{eqnarray}
with $\alpha_{k}=|\alpha_{k}|e^{i\varphi_{k}}$. Because the coupling
constant is a function of $\alpha_{k}$, according to the expressions
of $\alpha_{k}$ and $g_{k}'$ we can see that it is possible to
control this coupling constant by tuning the external fluxes
$\Phi_{xk}$ and controlling the evolution times $t_{k}$. Choosing
proper parameters such that $\theta=\pi/4$, the operator given by
Eq. (\ref{e12}) is equivalent to a nontrivial two-qubit gate
\cite{You1,You2,You3}.

\begin{figure}[htp]
\center
\includegraphics[width=8cm,height=3.6cm]{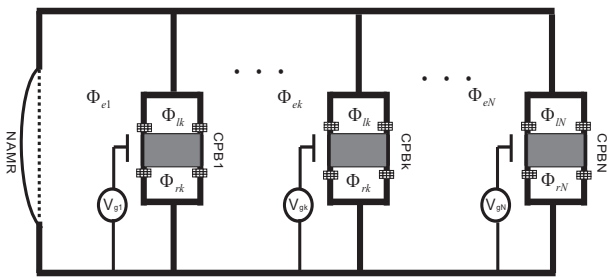}
\caption{ Schematic diagram of a multi-qubit superconducting
circuit in which $N$ qubits couple with an NAMR.}
\end{figure}

In practice, for performing a certain quantum computation mission,
we need to operate many qubits during the operation process. Hence
how to integrate the present model to multi-qubit circuit and how to
obtain selective and controllable two-qubit logic gates are main
missions. Fortunately, the integration of multi-qubit circuits is
straightforward. We plot the schematic diagram of our arrangement in
Fig. $2$. The coupling of the Josephson charge qubit $k$ with the
NAMR can be controlled by tuning the biasing fluxes $\Phi_{lk}$ and
$\Phi_{rk}$. For realizing a two-qubit logic gate between qubits $i$
and $j$, we couple the qubits $i$ and $j$ with the NAMR and decouple
other qubits with the NAMR by controlling the biasing fluxes.
Moreover, to obtain the required Hamiltonian given by Eq.
(\ref{e6}), we can control the biasing magnetic field threading
these loops to meet the condition $\sin(\pi\Phi_{bk}/\Phi_{0})=1$.

In conclusion, we have proposed a scheme to implement two-qubit
logic gates in two CPBs by coupling them with an NAMR. Under
different work conditions, we have shown that we can realize
two-qubit logic gates in three cases. In the first case, we
carefully tune the gate voltage and these biasing fluxes of the
CPBs to obtain controlled displacement operations between the CPBs
and the NAMR. And a certain sequence of controlled displacement
operations can reduce to a type of $\sigma_{x1}\sigma_{x2}$
interaction between the two CPBs. In the second case, we decouple
the coupling between the CPBs and the NAMR and also obtain a type
of $\sigma_{x1}\sigma_{x2}$ interaction between the two CPBs at
some selective time points. In the third method, we tune carefully
the energy separations of the CPBs such that the NAMR is large
detuning from the two CPBs. Under the large detuning condition, we
eliminate adiabatically the NAMR and obtain a type of
$(\sigma_{+1}\sigma_{-2}+\sigma_{-1}\sigma_{+2})$ interaction
between the two CPBs.

Finally, we give an estimation of the experimental feasibility for
the present scheme. In our scheme, we should consider the
following two time scales: the time required for implementing a
two-qubit logic gate and the lifetime of the qubits. We set the
following parameters \cite{Zho}, $B\approx0.1$ T, $l\approx30$
$\mu$m, $x_{0}\approx5\times10^{-13}$ m, $E_{J}^{0}\approx5$ GHz.
So we can obtain the maximum coupling constant $g'_{max}\approx30$
MHz. We choose a type of NAMR, $\omega\approx2\pi\times100$ MHz,
and quality factor $Q\approx10^{5}$. For simplicity, we assume
that the two qubits have the same parameters. In the first case,
for meeting the condition
$|\alpha_{1}||\alpha_{2}|\sin(\varphi_{2}-\varphi_{1})=\pi/4$, we
need to control the time $t_{k}$ of interaction between the qubit
$k$ and the NAMR to meet the equation $4g'_{1}g'_{2}\sin(\omega
t_{1}/2)\sin(\omega t_{2}/2)\sin(\omega
(t_{1}-t_{2})/2)/\omega^{2}=\pi/4$. Using the above parameters,
this equation reduces to $\sin(\omega t_{1}/2)\sin(\omega
t_{2}/2)\sin(\omega (t_{1}-t_{2})/2)=0.69$, so the fastest gate
implementation needs $t_{tot}\sim10^{-7}$ s. In the second case,
$\theta=4n\pi g'_{1}g'_{2}/\omega^{2}\sim\pi/4$ corresponds to
$t\sim10^{-7}$ s. Corresponding to the third case, we set the
detuning $\Delta\approx5g$, then the time required for
implementing a $\sqrt{i\texttt{SWAP}}$ gate
$t=\Delta\pi/(4g^{2})\approx1\times10^{-7}$ s. Moreover, for a
type of CPB, the dissipation time $T_{1}\approx1\sim10$ $\mu$s and
the dephasing time $T_{2}\approx0.1\sim1$ $\mu$s \cite{You}.
Therefore, in our present scheme the time required for
implementing two-qubit logic gates is shorter than the lifetime of
the qubit.

\acknowledgments

This work is supported by the National Natural Science Foundation
under Grant Nos. 10325523 and 10775048, and the National
Fundamental Research Program Grant No. 2007CB925204.

\end{document}